\documentclass[10pt,aps,twocolumn,prc,superscriptaddress,showpacs,nofootinbib,noshowkeys,floatfix,preprintnumbers]{revtex4-2}
\usepackage[dvipdfmx]{graphicx}
\usepackage[usenames]{color}
\usepackage{amsmath,amssymb}
\usepackage{multirow}
\usepackage{longtable}
\usepackage[normalem]{ulem}
\usepackage{epstopdf}
\usepackage{times}
\usepackage[normalem]{ulem}  % \sout{old text} for strikeout

\renewcommand\sout{\bgroup \color{red} \ULdepth=-.5ex \ULset}

\def\XXint#1#2#3{{\setbox0=\hbox{$#1{#2#3}{\int}$}
     \vcenter{\hbox{$#2#3$}}\kern-.5\wd0}}

\usepackage{bm}
\begin{document}
\title{Implications of latest NICER data for the neutron star equation of state}
\date{\today}
\author{Len Brandes}
\email{len.brandes@tum.de}
\affiliation{Technical University of Munich,  School of Natural Sciences,  Physics Department, 85747 Garching, Germany}
\author{Wolfram Weise}
\email{weise@tum.de}
\affiliation{Technical University of Munich,  School of Natural Sciences,  Physics Department, 85747 Garching, Germany}
\affiliation{Excellence Cluster ORIGINS, Boltzmannstr. 2, 85748 Garching, Germany}

\begin{abstract}
As an update to our previously performed Bayesian inference analyses of the neutron star matter equation-of-state and related quantities,  the additional impact of the recently published NICER data of PSR J0437-4751 is examined. 
Including the mass and radius distributions of this pulsar in our data base results in modest shifts from previously inferred median posterior values of radii $R$ and central densities $n_c$ for representative $1.4\,M_\odot$ and $2.1\,M_\odot$ neutron stars: radii are reduced by about $0.2-0.3$ km to values of $R_{1.4} = 12.1\pm 0.5$ km and $R_{2.1} = 11.9^{+0.5}_{-0.6}$ km (at the 68\% level), and central densities increase slightly to values of $n_c(1.4\,M_\odot)/n_0 = 2.8\pm 0.3$ and $n_c(2.1\,M_\odot)/n_0 = 3.8_{-0.7}^{+0.6}$ (in units of equilibrium nuclear matter density, $n_0 = 0.16$ fm$^{-3}$),  i.e.,  they still fall below five times nuclear saturation density at the 68\% level.  
As a further significant result,  the evidence established by analyzing Bayes factors  for a {\it negative} trace anomaly measure,  $\Delta = 1/3-P/\varepsilon < 0$,  inside heavy neutron stars is raised to {\it strong}. 
\end{abstract}
\pacs{}

\maketitle

\section{Introduction}
Determinations of the mass and the radius of the nearest and brightest millisecond pulsar,  PSR J0437-4751,  have recently been reported by the NICER collaboration \cite{Choudhury2024}.  Their analysis benefits from previous accurate radio measurements of orbital and astrometric parameters and a tightly constrained distance of the object \cite{Reardon2024}.  Among the reported values,
\begin{eqnarray}
	M &=& 1.418\pm 0.037\,M_\odot~, \\
	R &=& 11.36^{+0.95}_{-0.63}\,\text{km}~,
\end{eqnarray}
the radius is seen to be tendentially smaller in comparison with a previously measured pulsar of similar mass, PSR J0030+0451 with mass $M= 1.34^{+0.15}_{-0.16}\,M_\odot$ and radius $R=12.71^{+1.14}_{-1.19}$ km \cite{Riley2019}.  The question therefore emerges as to how this new information, when added to the existing database,  influences the inferred constraints on the equation of state (EoS) of neutron star matter.  A first analysis in this context has been performed in Ref.\,\cite{Rutherford2024}.  

At the same time, the radius of the heavy neutron star PSR J0740+6620 with a mass of $M = 2.07\pm 0.07\,M_\odot$ has been updated from Ref. \cite{Riley2021} and slightly increased to $R = 12.49^{+1.28}_{-0.88}$ km \cite{Salmi2022,Salmi2024}.  While this marginal shift within the uncertainty band is not expected to lead to any significant change in the inference results,  we include it nonetheless in our data set.

The present paper examines how these new pieces of information,  in particular the J0437-4751 data,  add to our previous inference analysis \cite{Brandes2023a}.  It also offers insights as to how new observations and measurements can contribute to updating results for the inferred equation of state (EoS). 

\section{Inference procedure}

Numerous Bayesian inference analyses have been carried out in the literature (see, e.g., \cite{Marczenko2023,Annala2023,Han2023,Essick2023,Takatsy2023,Jiang2023,Brandes2023,Brandes2023a,Mroczek2023b,Pang2023,Lim2024,Tsang2023,Fan2024,Koehn2024,Ofengeim2024,Komoltsev2024,Malik2024,Tang2024} for a selection of recent works), with a growing interest in complementary approaches using machine learning methods 
\cite{Soma2023,Krastev2023,Guo2023,Chatterjee2023,Zhou2023,Carvalho2023,Carvalho2024,Fujimoto2024,Farrell2023a,Farrell2023b,Brandes2024a}. In this work, we employ the same inference framework as in \cite{Brandes2023a}.  The prior parameter space is prepared in the form of general segment-wise linear interpolations of the squared speed of sound,  $c_s^2(\varepsilon) = {\partial P(\varepsilon)\over\partial\varepsilon}$, where $P$ denotes the pressure and  $\varepsilon$ denotes the  energy density.  Previous systematic tests \cite{Brandes2023} have confirmed that for this choice of parametrization the inferred posterior results are driven by the empirical data sets and not by the choice of prior,  an important prerequisite for achieving  reliable constraints on the EoS with controlled uncertainty bands. 

Our procedure differs from the one reported in \cite{Rutherford2024} in several respects.  First,  the equation of state models used in \cite{Rutherford2024} consist of a three-segments polytropic parametrization of the pressure (PP),  or a single Gaussian form of the squared speed of sound with asymptotic conformal behavior imposed (CS).  The PP or CS choices are then shown to result in significantly different posterior pressure distributions,  indicating that one or both of these prior parametrizations may be too restrictive.  In this context, see also the discussion in \cite{Legred2022} on the limited capability of (un-skewed) Gaussian parametrizations of the sound speed.  Our ansatz for the squared sound speed prior \cite{Brandes2023a},
\begin{eqnarray}
c_{s}^2(\varepsilon,\theta) = \frac{(\varepsilon_{i+1} - \varepsilon)c_{s,i}^2 +(\varepsilon - \varepsilon_i)c_{s,i+1}^2}  {\varepsilon_{i+1} - \varepsilon_i}~,
\end{eqnarray}
in terms of $i = 1,\dots,N$ segments and employing a broadly varying parameter set,  $\theta = \{c_{s,i}^2,\varepsilon_i\}$,  permits a wide sampling of input into the inference procedure.  It leads to posterior output results that are not affected by changes of the prior if a sufficiently large number $N$ of segments is used to start with.  In practice it has been tested that the posterior results remain stable for $N\gtrsim 4-5$ \cite{Altiparmak2022} and compare well with the results of a non-parametric Gaussian process \cite{Annala2023}.  Here we choose an even larger number of segments,  namely $N=6$.

Perhaps even more important are the differences in the treatment of the chiral effective field theory (ChEFT) constraints at low densities.  In Ref.\,\cite{Rutherford2024} these constraints are imposed as {\it priors}.  A matching of the ChEFT prior at baryon densities $1.1\,n_0$ or $1.5\,n_0$ (with $n_0 = 0.16$ fm$^{-3}$ the equilibrium density of isospin-symmetric nuclear matter) to the higher-density behavior is observed in \cite{Rutherford2024} to have a strong influence on the further evolution of the EoS beyond the matching point,  enforcing a steep rise of the pressure shortly above that matching point.  We argue instead \cite{Brandes2023,Brandes2023a} that within the Bayesian inference procedure the ChEFT constraint should be introduced as a {\it likelihood} rather than a prior,  analogous to the way the set of empirical astrophysical data is treated.  Indeed,  chiral EFT should properly be interpreted as an efficient parametrization representing large amounts of empirical nuclear physics data by fitting sets of low-energy constants in a controlled expansion.  In practice the ChEFT likelihood is employed conservatively up to $1.3\,n_0$ \cite{Essick2020}. An even more restrictive choice of $1.1\,n_0$ leads to comparable posterior results \cite{Brandes2023a}. A similar likelihood treatment of the low-density constraint has been performed in \cite{Koehn2024}.

At asymptotically high energy densities $\varepsilon$,  far above the values accessible in the cores of even the heaviest neutron stars,  the perturbative QCD (pQCD) limit for the EoS is implied.  In particular the speed of sound reaches the conformal bound,  $c_s = \sqrt{1/3}$.  Recent results for the pressure derived from pQCD at next-to-next-to-next-to-leading order (N3LO) \cite{Gorda2023a} indicate an applicability range of perturbative QCD that starts upward from baryon chemical potentials $\mu_B \gtrsim 2.4$ GeV,  or baryon densities $n_B \gtrsim 40\,n_0$.  A corresponding likelihood that measures whether a given EoS can be connected to the pQCD results in a causally and thermodynamically stable way,  is implemented following \cite{Komoltsev2022,Gorda2023}.  As in  Ref. \cite{Brandes2023a} (but in contrast to some other recent analyses) each EoS for the computation of the posterior credible bands is employed only up until its respective maximum central energy density $\varepsilon_{c,\max}$ corresponding to the last stable neutron star in the mass-radius sequence.  

\section{Results}

The present inference analysis follows the steps already described in \cite{Brandes2023a}.  Starting from prior samples of $c_s^2$ defined by the chosen distribution of parameters $\{c_{s,i}^2, \varepsilon_i\}$, the posterior sound speed distribution based on the empirical data and constraints can then be computed using Bayes' theorem.  Table \ref{tab:DataSet} summarizes the input information used in the present analysis.  The set denoted `Standard' refers to previous data from Shapiro delay \cite{Arzoumanian2018,Antoniadis2013},  NICER \cite{Riley2021,Salmi2022,Salmi2024,Riley2019} and gravitational wave measurements \cite{Abbott2019,Abbott2020}.

\begin{table*}[!htb]
	\centering \small
	\begin{tabular}{lllll}
		\hline \hline
		&\multicolumn{3}{l}{Data and constraints}&  \\ \hline
		&& PSR J1614–2230 & $M = 1.908 \pm 0.016 \, M_\odot$ \cite{Arzoumanian2018} & \qquad \\
		&& PSR J0348+0432 & $M = 2.01 \pm 0.04 \, M_\odot$ \quad \cite{Antoniadis2013} & \\
		&& PSR J0740+6620 \qquad \qquad  & $ M = 2.073 \pm 0.069 \, M_\odot$&  \\
		&Standard && $R = 12.49^{+1.28}_{-0.88}\,\text{km}$ \quad \cite{Riley2021,Salmi2022,Salmi2024}& \\[.8ex]
           	&& PSR J0030+0451 & $M=1.34^{+0.15}_{-0.16} \, M_\odot$ & \\
		&&& $R =12.71^{+1.14}_{-1.19}\,\text{km}$ \quad \cite{Riley2019}& \\[.8ex]
		&& GW170817 & $\bar{\Lambda}=300^{+420}_{-230}$ \quad \cite{Abbott2019}& \\
		&& GW190425 & $\bar{\Lambda} \leq 600$ \quad \cite{Abbott2020}& \\ [.8ex]\hline
		&BW & \multicolumn{1}{l}{PSR J0952-0607} & $M = 2.35 \pm 0.17 \, M_\odot$  \cite{Romani2022} & \\ [.8ex]  
		& J0437 & \multicolumn{1}{l}{PSR J0437-4715} & $M=1.418\pm0.037 \, M_\odot$ & \\
		&&& $R=11.36^{+0.95}_{-0.63}\,\text{km}$ \quad \cite{Choudhury2024} & \\[.8ex]\hline 
		&& ChEFT &$(n_B\leq 1.3\,n_0)$\quad \cite{Drischler2021,Drischler2022}& \\
		&& pQCD &$(n_B\gtrsim 40\,n_0)$\quad \cite{Gorda2021,Komoltsev2022,Gorda2023,Gorda2023a}& \\ [.8ex]  
		\hline \hline  
	\end{tabular}
	\caption{\small \small Data and constraints used in the Bayesian inference analysis. The `Standard' set includes data from Shapiro delay measurements,  from earlier NICER observations,  and from gravitational wave signals of binary neutron star mergers. We examine the additional influence of the new NICER data (J0437) together with the impact of the heavy `black widow' pulsar J0952-0607 (BW).  All listed data are at the 68\% level except for the tidal deformabilities ($\bar{\Lambda}$) based on the gravitational wave data,  which are at the 90\% level.  Chiral effective field theory (ChEFT) and perturbative QCD (pQCD) constraints at low and asymptotic baryon densities,  $n_B$,  are implemented as described in the text.}
	\label{tab:DataSet}
\end{table*}

Given $c_s^2(\varepsilon)$ the EoS distributions are evaluated as  
\begin{eqnarray}
P(\varepsilon)=\int_0^\varepsilon d\varepsilon'\,c_s^2(\varepsilon')~.
\end{eqnarray}
The derivation of the posterior distribution involves multiple solutions of the Tolman-Oppenheimer-Volkoff equations together with the differential equation determining the tidal deformabilities \cite{Hinderer2010},  both of which require $P(\varepsilon)$ as input.  Further quantities of interest are derived solving the Gibbs-Duhem equation at zero temperature,  $P+\varepsilon = \mu_B\,n_B$,  with the baryon density and baryon chemical potential given as:
\begin{eqnarray}
n_B = \frac{\partial P}{\partial\mu_B}~~,\quad \mu_B=\frac{\partial\varepsilon}{\partial n_B}=\frac{P+\varepsilon}{n_B}~.
\end{eqnarray}
In the following, we summarize and discuss inference results with inclusion of the new NICER data (J0437) compared to those of previous work.  It is also of interest to present selected results with and without the heavy `black widow' (BW) pulsar J0952-0607 \cite{Romani2022} in order to examine its influence.  This is one of the fastest rotating compact objects,  so the measured BW mass requires rotational corrections.  These radius-dependent corrections have been applied as in \cite{Brandes2023a} based on the empirical formula derived in \cite{Konstantinou2022},  reducing the observed BW mass ($2.35\pm 0.17\,M_\odot$) to an equivalent non-rotating mass (e.g.  to a value of approximately $2.3\pm 0.2\,M_\odot$ at a radius of about 12 km). 

\begin{figure*}[!htb]
	\centering
	\includegraphics[height=50mm,angle=-00]{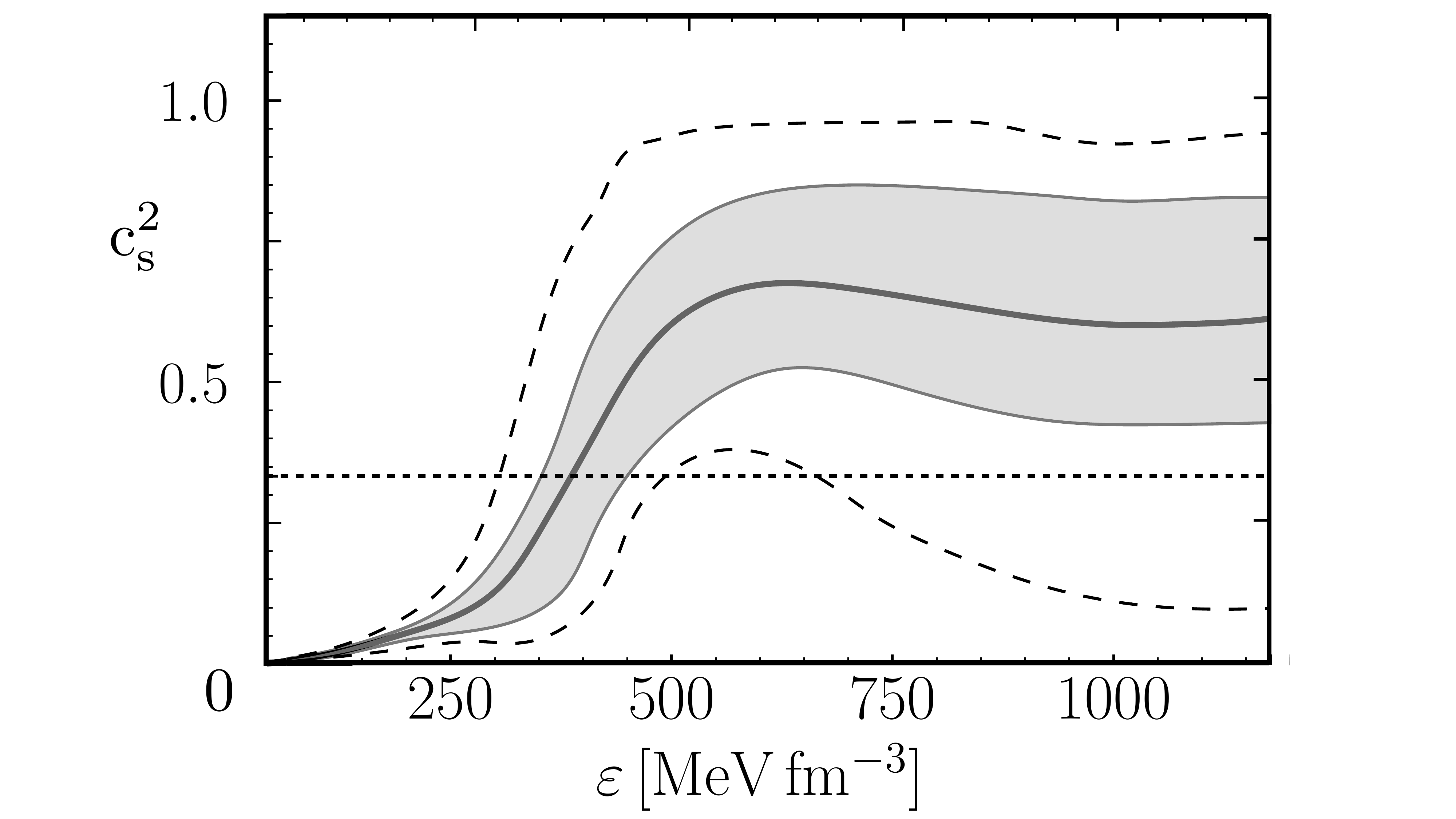} 
        \includegraphics[height=50mm,angle=-00]{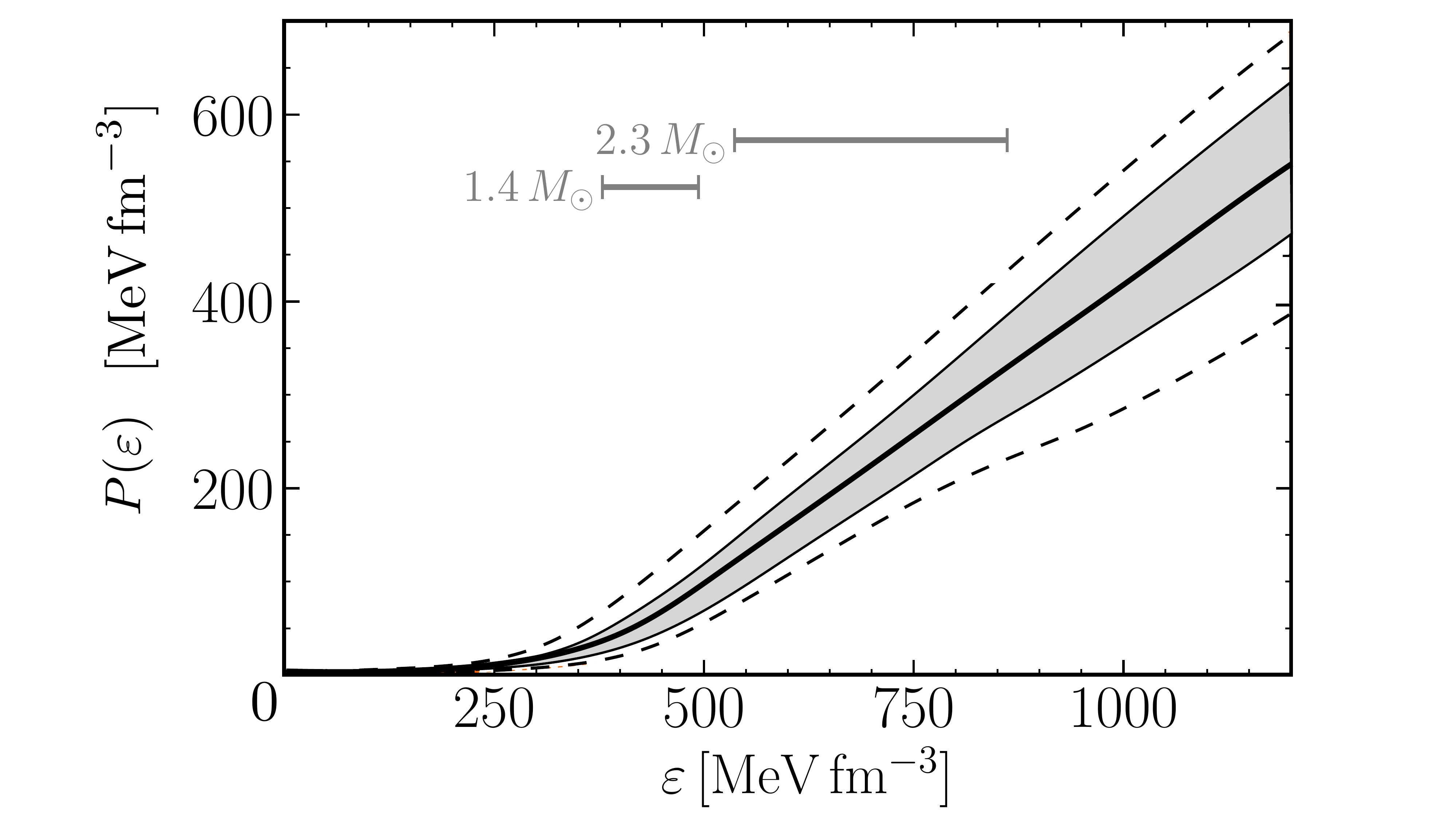} \\
	\caption{\small Marginal 95\% (dashed lines) and 68\% (gray) posterior credible bands and medians (solid lines) for the squared speed of sound, $c_s^2(\varepsilon)$, and the pressure, $P(\varepsilon)$,  as a function of energy density $\varepsilon$.  The underlying empirical data sets and constraints are listed in Table\,\ref{tab:DataSet} and include the new NICER data (J0437)  as well as the mass of the heaviest pulsar (BW) after rotational correction.  The dotted line in the left figure marks the conformal bound,  $c_s^2 = 1/3$.  In the EoS figure (right) the central energy density ranges in the cores of neutron stars with masses 1.4 and 2.3 $M_\odot$ are indicated. }
	\label{fig:figure1}
\end{figure*}

\begin{figure*}[!htb]
	\centering
	\includegraphics[height=50mm,angle=-00]{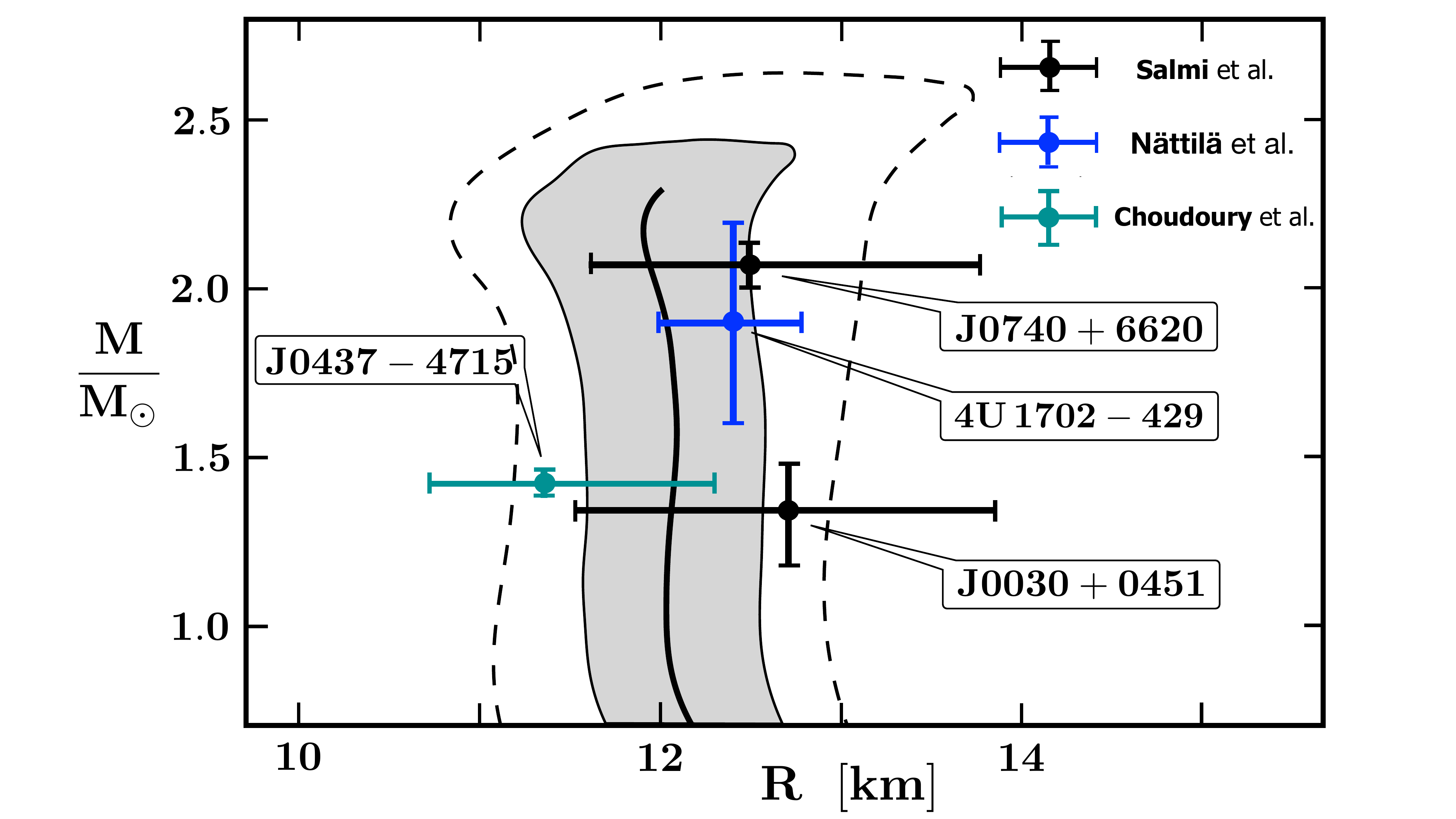} 
	\includegraphics[height=50mm,angle=-00]{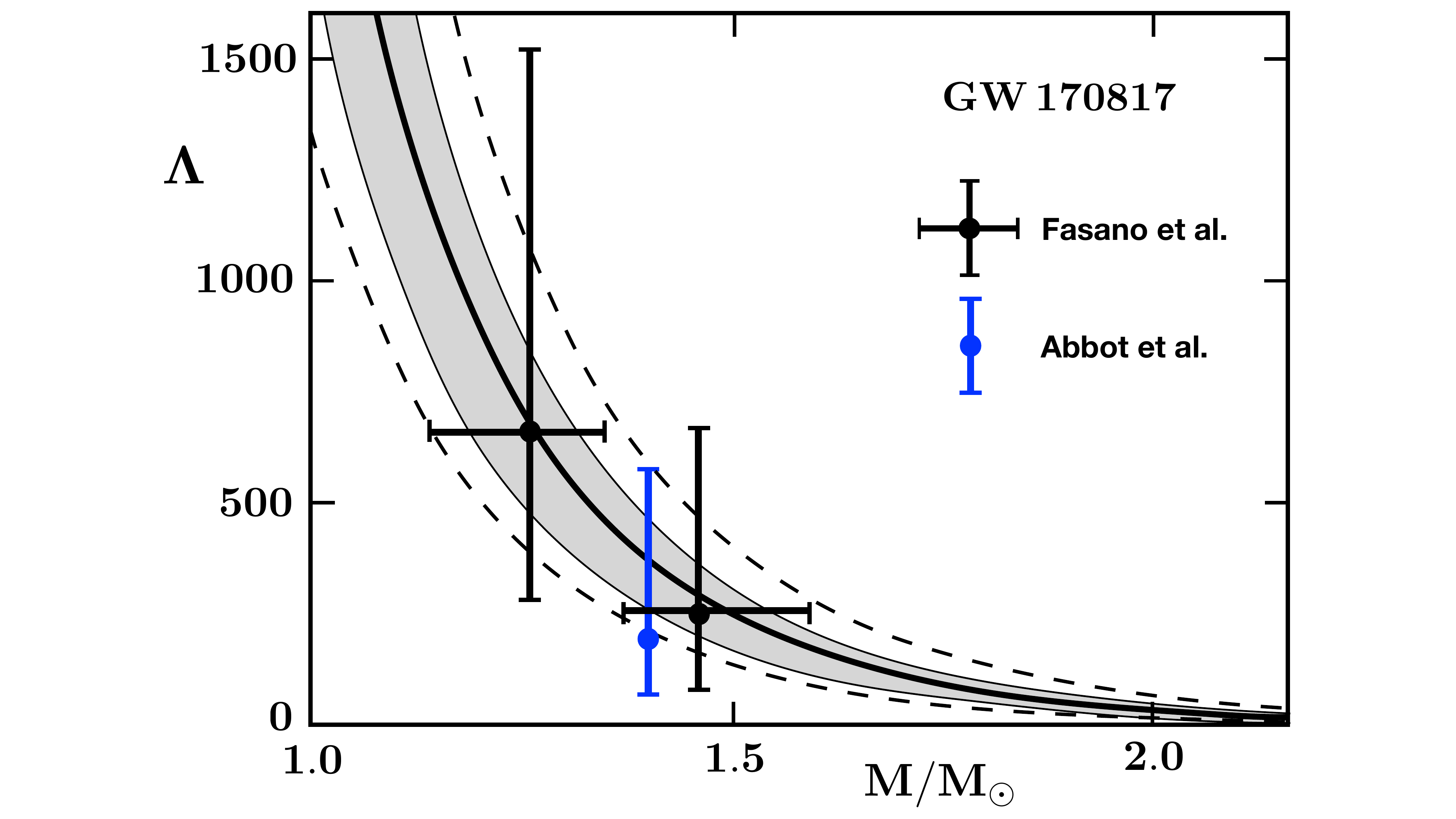} \\
	\caption{\small Marginal 95\% and 68\% posterior credible bands and medians for the mass-radius relation (left)  and tidal deformability as a function of neutron star mass $M$ (right),  inferred from all data listed in Tab.\,\ref{tab:DataSet}.  The mass-radius relation is compared to the marginalized intervals at the 68\% level from the NICER data analyses of PSR J0030+0451 and PSR J0740+6620 \cite{Riley2019,Riley2021,Salmi2022,Salmi2024},  as well as PSR J0437-4715 \cite{Choudhury2024}.  In addition the 68\% mass-radius credible intervals of the thermonuclear burster 4U 1702-429 \cite{Naettilae2017} is displayed which is not included in the Bayesian analysis. $\Lambda(M)$ is compared to the masses and tidal deformabilities inferred in Ref.\,\,\cite{Fasano2019} for the two neutron stars in the merger event GW170817 at the 90\% level (black) as well as $\Lambda(1.4\,M_\odot)$ at the 90\% level extracted from GW170817  \cite{Abbott2018} (blue).}
	\label{fig:figure2}
\end{figure*}

\begin{figure*}[!htb]
	\centering
		\includegraphics[height=49mm,angle=-00]{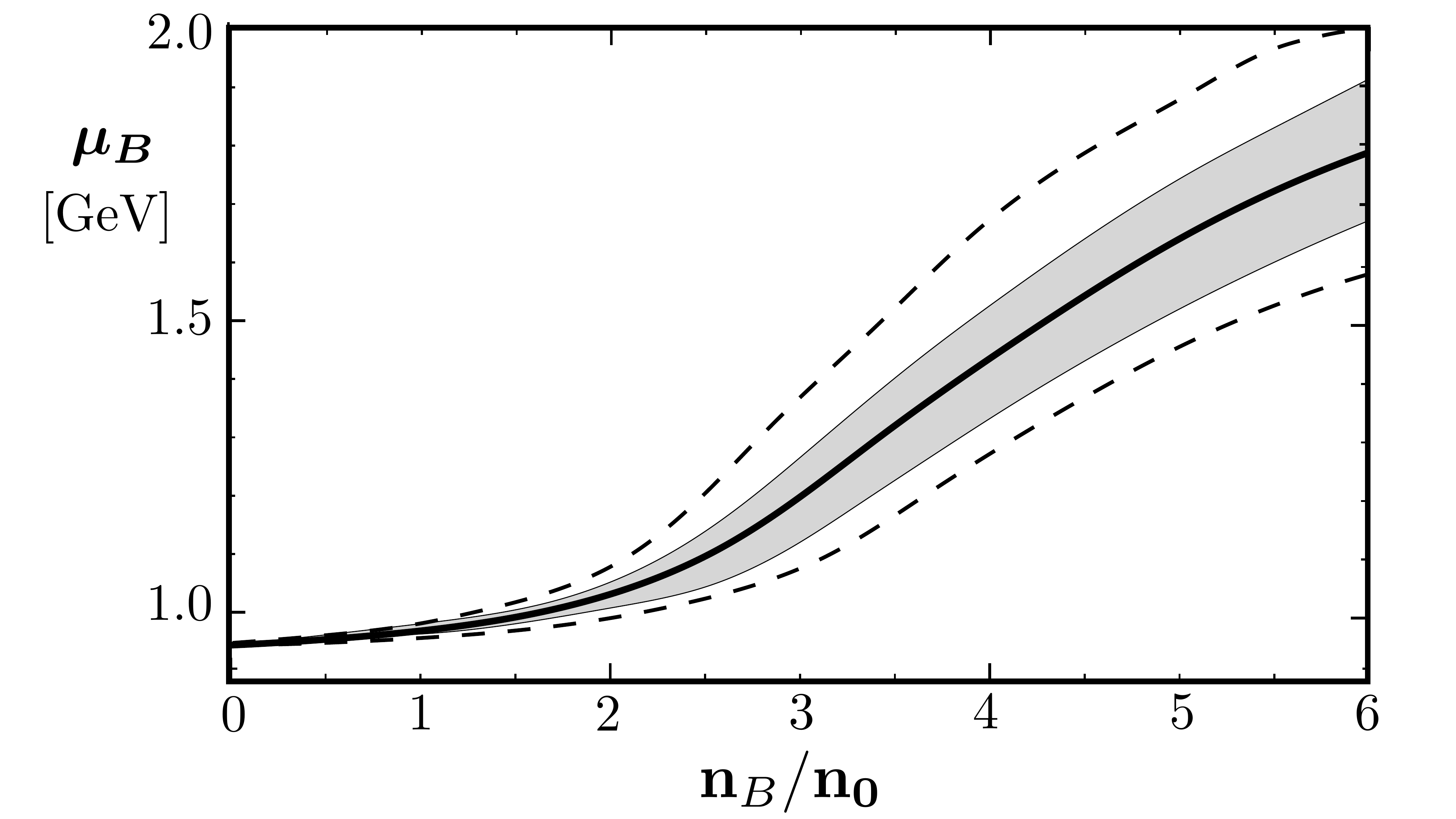}
 		\includegraphics[height=48mm,angle=-00]{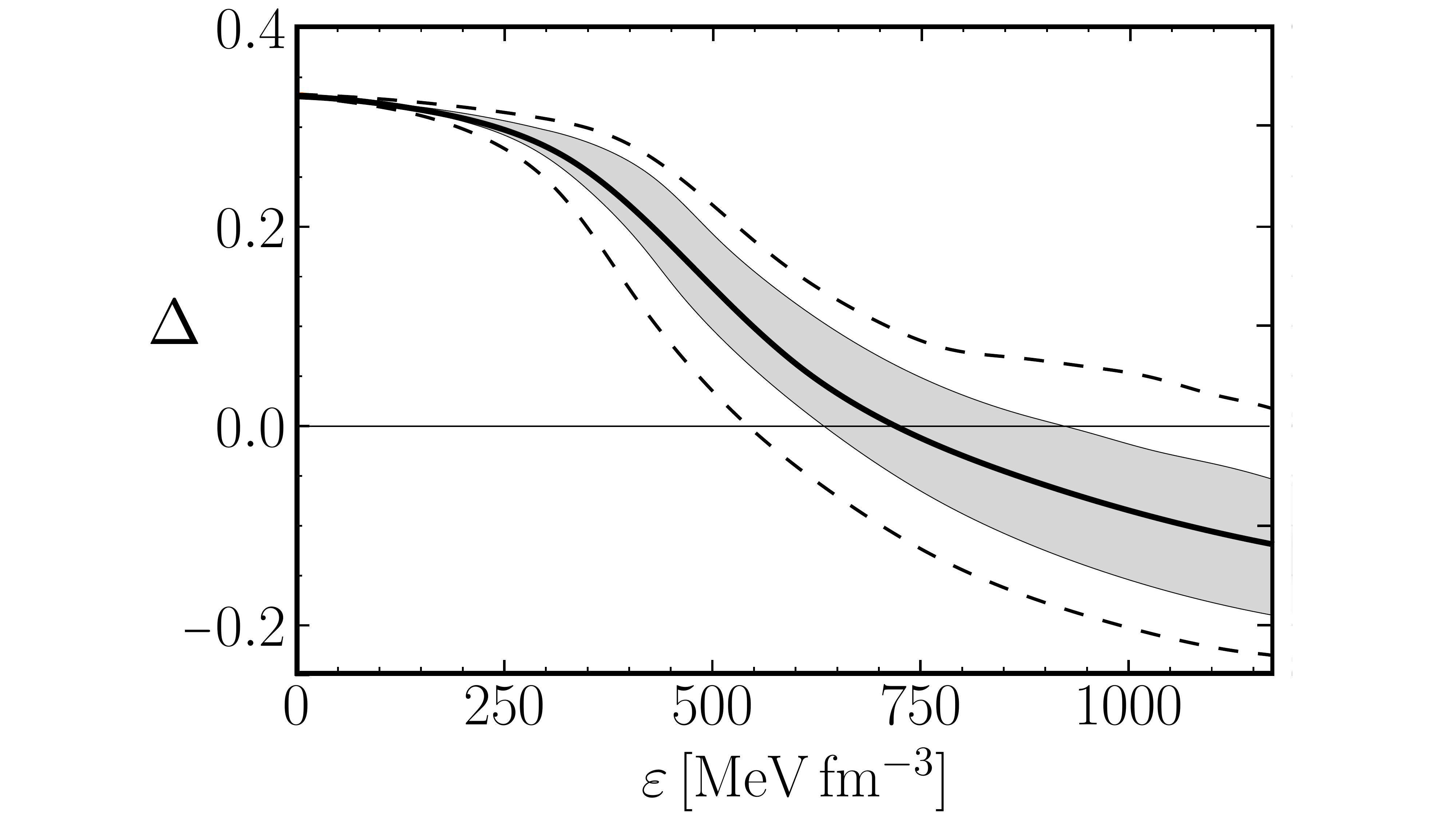}		
		\caption{\small Posterior 95\% (dashed) and 68\% (gray) credible bands and medians (solid lines) for baryon chemical potential $\mu_B$ as function of baryon density (left),  and the trace anomaly measure, $\Delta = 1/3 - P/\varepsilon$,  as function of energy density $\varepsilon$ (right). }
		\label{fig:figure3}
\end{figure*}
The posterior credible bands of the squared sound speed and equation of state,  including the information both from PSR J0952-0607 and PSR J0437-4715,  are shown in Fig.\,\ref{fig:figure1}.  As already noted in \cite{Brandes2023a},  the inferred $P(\varepsilon)$ turns out to be even stiffer than the time-honored Akmal–Pandharipande–Ravenhall (APR) equation of state \cite{Akmal1998} throughout the region of energy densities between 0.4 and 0.9 GeV/fm$^3$ relevant for neutron stars.  Posterior credible bands of the mass-radius relation and the tidal deformability in comparison with empirical data are displayed in Fig.\,\ref{fig:figure2}.  The influence of the new NICER (J0437) data can be examined by comparing with the corresponding results in Refs. \cite{Brandes2023a, Brandes2024} that do not include J0437. In comparison, the median of the $M-R$ plot is now shifted to slightly smaller radii of about 12 km with a $68\%$ credible band of about $\pm 0.5$\,km.  The updated EoS at low densities shows a tendency to be somewhat softer,  implying a stiffer sound speed at higher densities. 

The J0437 radius might seem to be in mild tension with the larger radius of PSR J0030+0451 (though with overlapping 68\% uncertainty bands). But the general qualitative pattern of the inferred mass-radius relation remains quite similar after the inclusion of J0437.  In particular,  the squared speed of sound exceeds the conformal bound, $c_s^2 = 1/3$,  at baryon densities $n_B\simeq 2-3\,n_0$ and stays well above this bound throughout the range of energy densities realized in the cores of neutron stars.  

Table \ref{tab:NS_properties} summarizes inferred characteristic properties of 1.4 and 2.1 solar mass neutron stars when the J0437 information is included in addition to the `Standard' data set consisting of the GW, Shapiro and previous NICER data together with the constraints from ChEFT and pQCD listed in Tab. \,\ref{tab:DataSet}.  Table \ref{tab:NS_properties} also outlines the additional impact of the heavy BW mass on the inferred properties of typical neutron stars.  In comparison with previous results,  the relatively small radius of PSR J0437-4715 reported in \cite{Choudhury2024} has the effect of shifting the $R$-median downward by about $0.2 - 0.3$ km to values close to $12\,$km,  still well within the inferred 68\% posterior credible bands.  This trend was already reported in the preliminary analyses \cite{Malik2024,Tang2024} and is accompanied by a modest increase by less than $8\%$ of the central densities reached in the cores of neutron stars.  The central density of a 2.1 $M_\odot$ neutron star remains below five times $n_0$ (at the 68\% level).  The inferred properties of a possible ultra-heavy 2.3 $M_\odot$ star differ only marginally,  with medians and 68\% posterior credible intervals of central density, $n_c/n_0 = 4.0^{+0.6}_{-0.9}$,  and radius $R=12.0\pm0.6$ km.  The maximum supported mass shifts from $M_{\max} = 2.22_{-0.15}^{+0.10}\,M_\odot$ (Standard + J0437 data) to $M_{\max} = 2.30_{-0.15}^{+0.12}\,M_\odot$ with inclusion of the additional heavy black-widow mass data.

\begin{table*}[!htb]
	\centering \small
	\begin{tabular}{lllllllllll}
		\hline \hline 
		&\multicolumn{2}{l}{} & \multicolumn{3}{c}{Standard + J0437} &\qquad\qquad& \multicolumn{3}{c}{Standard + J0437 + BW } &\\ \cline{4-6} \cline{8-10}
		&\multicolumn{2}{l}{} && \multicolumn{1}{c}{95\%} & \multicolumn{1}{l}{68\%} &&& \multicolumn{1}{c}{95\%} & \multicolumn{1}{l}{68\%} & \\ \hline
		&&$n_c /n_0$ && $2.8\pm0.6$ & $_{-0.3}^{+0.4}$ &&& $2.8\pm0.6$ &$\pm0.3$ &\\ [.8ex]
		&&$\varepsilon_c  \, $[MeV$\,$fm$^{-3}$] \qquad \qquad & \qquad & $459_{-107}^{+109}$ & $_{-50}^{+62}$ && \quad \qquad & $444_{-111}^{+99}$ & $_{-64}^{+50}$ &\\ [.8ex]
		&$1.4\, M_\odot$& $P_c \, $[MeV$\,$fm$^{-3}$] && $68_{-23}^{+24}$ & $_{-14}^{+11}$ &&& $65_{-21}^{+24}$ & $_{-15}^{+10}$ & \\ [.8ex]
		&&$R \, $[km] && $12.0\pm0.9$ & $\pm 0.5$ &&& $12.1_{-0.8}^{+0.9}$ & $\pm 0.5$ &\\ [.8ex]
		&&$\Lambda$ && $364_{-155}^{+204}$ & $_{-122}^{+72}$ &&& $375_{-156}^{+215}$ & $_{-114}^{+86}$ &\\ [.8ex] \hline
		&&$n_c /n_0$ && $4.1_{-1.4}^{+1.6}$ & $_{-0.7}^{+0.8}$ &&& $3.8_{-1.2}^{+1.4}$ & $_{-0.7}^{+0.6}$ &\\ [.8ex]
		&&$\varepsilon_c  \, $[MeV$\,$fm$^{-3}$] \qquad \qquad & \qquad & $729_{-313}^{+348}$ & $_{-197}^{+130}$ && \qquad & $659_{-243}^{+314}$ & $_{-144}^{+125}$ &\\ [.8ex]
		&$2.1\, M_\odot$& $P_c \, $[MeV$\,$fm$^{-3}$] && $235_{-135}^{+192}$ & $_{-101}^{+51}$ &&& $204_{-111}^{+153}$ & $_{-67}^{+60}$ &\\ [.8ex]
		&&$R \, $[km] && $11.7_{-1.0}^{+1.2}$ & $_{-0.6}^{+0.5}$ &&& $11.9_{-1.0}^{+1.2}$ & $_{-0.6}^{+0.5}$ &\\ [.8ex]
		&&$\Lambda$ && $19_{-13}^{+24}$ & $_{-10}^{+6}$ &&& $23_{-15}^{+26}$ & $_{-11}^{+8}$ &\\ [.8ex]
		\hline \hline 
	\end{tabular}
	%\end{ruledtabular}
	\caption{\small Median, 95\% and 68\% credible intervals for selected neutron star properties based on the `Standard' data set from Tab.\,\ref{tab:DataSet} and the new NICER data of PSR J0437-4715. These properties are computed from the one-dimensional posterior probability distributions marginalized over all other parameters.  Listed are the central density,  central energy density,  central pressure,  radius and tidal deformability of neutron stars with masses $M = 1.4\,M_\odot$ and $M = 2.1\,M_\odot$.  The two right columns  include the additional information from the black widow (BW) pulsar PSR J0952-0607.}
	\label{tab:NS_properties}
\end{table*}

The PSR J0030+0451 case requires further discussion as it has been reevaluated in 
Ref. \,\cite{Vinciguerra2024}. The updated in-depth analysis reveals a more complicated structure of PSR J0030+0451 than previously anticipated in \cite{Riley2019}.  Two models,  both compatible with the NICER and XMM-Newton data,  arrive at different mass-radius values that are barely compatible with each other ($M=1.40^{+0.13}_{-0.12}\,M_\odot$ and $R=11.71^{+0.88}_{-0.83}$ km,   as compared to $M=1.70^{+0.18}_{-0.19}\,M_\odot$ and $R=14.44^{+0.88}_{-1.05}$ km \cite{Vinciguerra2024}).  Given this ambiguous situation we have also carried out an inference computation with the uncertain J0030 data removed from the analysis. The relative statistical weight of the J0437 data in the 1.4 solar mass sector is thus increased.  The changes in the resulting posterior distributions are nevertheless minor: the radii at all neutron star masses shift downward by only about 0.1 km since the larger radius of PSR J0740+6620 ($R\simeq 12.5$ km) must still be supported.  Central densities of neutron star cores move upward correspondingly by small amounts well within the uncertainty bands.

Recently the NICER collaboration also published the analysis of a fourth pulsar,  PSR J1231-1411 \cite{Salmi2024a}.  However,  their inference procedure converged only for restrictive radius priors.  For example,  when the radius prior was chosen to be consistent with previous observational constraints by employing the posterior results of the Bayesian analysis in \cite{Raaijmakers2021},  mass and radius values of $M = 1.04_{-0.03}^{+0.05}\,M_\odot$ and $12.6\pm0.3\,$km were inferred.  These results are consistent with the 68\% posterior credible displayed in Fig.\,\,\ref{fig:figure2}.  When an uninformed radius prior between zero and 14$\,$km was employed instead,  the authors reported a much larger radius of $13.5_{-0.5}^{+0.3}\,$km,  in tension with the posterior results inferred here as well as in other recent analyses \cite{Koehn2024,Malik2024,Tang2024}.  In view of this ambiguity we have chosen not to include PSR J1231-1411 in our data base.

Another recently reported source,  HESS J1731-347,  suggests a small radius, $R \sim 10\,$km, in combination with an unusually small mass,  $M \sim 0.77\,M_\odot$ \cite{Doroshenko2022}. This object has large systematic uncertainties associated with varying assumptions about atmosphere models (see e.g.  the discussion in \cite{Koehn2024}).  Adding it nonetheless to our data base leads to changes similar to those discussed in \cite{Brandes2023a}: the mass-radius relation is moved towards even lower radii,  requiring a more rapid increase of the sound speed.   The inclusion of J0437 already shifts the mass-radius credible band to smaller radii so that it now overlaps with the upper limt of the HESS J1731-347 radius data at the 68\% level,  unlike our previous results without those latest NICER data.

Further instructive information can be extracted by examining the inferred baryon chemical potential, $\mu_B$,  and the trace anomaly measure (with $T_\mu^\mu$ the trace of the QCD energy momentum tensor): 
\begin{eqnarray}
\Delta = \frac{T_\mu^\mu}{3\,\varepsilon} = \frac{1}{3}-\frac{P(\varepsilon)}{\varepsilon}~,
\label{eq:delta}
\end{eqnarray}
both displayed in Fig.\,\ref{fig:figure3}.  The baryon chemical potential shows a characteristic strong increase at baryon densities $n_B \simeq 2-4\,n_0$,  indicating a rapidly growing strength of repulsive correlations in the transitional range between nuclear and neutron star matter.  The trace anomaly measure (\ref{eq:delta}) is of special interest as it has been subject of discussions about the possible approach to conformal matter, $\Delta\rightarrow 0$, within the core region of heavy neutron stars \cite{Fujimoto2022}.  The data-driven inferred $\Delta$ with inclusion of J0437 features a strong tendency towards negative values of $\Delta$,  i.e.,  pressure $P$ exceeding $\varepsilon/3$,  at baryon densities $n_B > 3\,n_0$.  In fact, using the data set with inclusion of J0437,  the computation of Bayes factors comparing the likelihoods for $\Delta < 0$ against $\Delta \geq 0$ yields:
\begin{eqnarray}
{\cal B}_{\Delta \geq 0}^{\Delta < 0} &=& 5.6~~~~~~\dots\text{without BW}~, \nonumber \\
 &=& 10.9~~~~\dots\text{with BW}~.
\label{eq:bayes}
\end{eqnarray}
Accordingly, the evidence for a {\it negative} trace anomaly measure inside neutron stars is raised from moderate (as reported in \cite{Brandes2023a}) to {\it strong} when the new NICER data are included in the inference procedure in addition to the `black widow' pulsar.  An interesting recent discussion of phenomena that can lead to a negative trace anomaly measure at high density,  such as a superfluidity gap,  can be found in \cite{Fukushima2024}.

\section{Summary and Conclusions}

The inferred equation of state of neutron star matter has been updated by including the new NICER results of PSR J0437-4751. While the general pattern of previous analyses is maintained,  the following moderate changes of neutron star properties have emerged in comparison with the results obtained previously in Ref.\,\cite{Brandes2023a}. Notably the common data set on which this comparison is based includes PSR J0030+0451 (with a similar mass but larger radius than J0437).  Also included is the mass of the heavy black widow pulsar PSR J0952-0607: 

(i) At the 68\% level,  the radius of a 1.4 solar mass neutron star is reduced from $R_{1.4} = 12.3_{-0.5}^{+0.4}$ km to $R_{1.4} = 12.1\pm 0.5$ km with the inclusion of J0437\footnote{Note that the previous values slightly differ from those reported in \cite{Brandes2023a} due to the updated radius value of PSR J0740+6620 \cite{Salmi2024}}.  The radius of a 2.1 solar mass neutron star is similarly reduced from $R_{2.1}= 12.2^{+0.5}_{-0.7}$ km to $R_{2.1}=11.9^{+0.5}_{-0.6}$ km.  These systematic changes still fall within the inferred $68\%$ credible bands of the previous analysis.  An interesting feature is that the median of the neutron star radii remains stable at $R \simeq 12$ km,  almost independent of the neutron star mass, with a credible band of about $\pm 1$ km width at the 95\% level. Given the uncertainties of the reanalyzed PSR J0030+0451 data,  we have also examined the minor changes that occur when these data are omitted from the input set.

(ii) The slightly reduced neutron star radii are accompanied by slightly increased values of the central core densities. At the 68\% level the central baryon density $n_c/n_0$ (in units of the equilibrium nuclear matter density $n_0$) of a 1.4 $M_\odot$ star increases from $2.6 \pm 0.3$ to $2.8 \pm 0.3$ with the inclusion of J0437; for a 2.1 $M_\odot$ star $n_c/n_0$ increases from $3.6 \pm 0.7$ to $3.8^{+0.6}_{-0.7}$.  Notably these central density values are sensitive to the presence of the black-widow pulsar PSR J0952-0607 with $M\simeq 2.3\,M_\odot$ in the database.  Omitting this heavy pulsar would increase the inferred central density of a 2.1 $M_\odot$ neutron star by about 10\%.  In any case the median values of central baryon core densities remain well below $5\,n_0$ even for stars as heavy as $2.3\,M_\odot$.  In a baryonic picture of neutron star matter this implies average distances between baryons of 1 fm or larger, at which the hard cores of baryons are still well separated \cite{Brandes2024,Kaiser2024}.  

(iii) The baryon chemical potential with inclusion of J0437 differs only marginally from the previously inferred $\mu_B$ in \cite{Brandes2023a}.  Its strongly rising behavior at baryon densities $n_B \sim 2-4\,n_0$ can be interpreted as a signal of increasingly repulsive many-body forces.  In a relativistic Fermi liquid description with neutron-like quasiparticles \cite{Friman2019,  Brandes2024},  guided by the median of the inferred $\mu_B$,  the effective quasiparticle potential does indeed show a density dependence characteristic of strongly repulsive many-body correlations beyond two-body interactions.  In the updated inferred $\mu_B$ and EoS,  the previously established Bayes-factor based evidence \cite{Brandes2023a,Brandes2024} against a strong first-order phase transition in neutron stars with masses $M\lesssim 2.1\,M_\odot$ persists.  On the other hand, the occurrence of a continuous crossover from hadronic matter to some form of strongly correlated quark matter can well be realized.  For example,  the latest version of an EoS motivated by the idea of quark-hadron continuity,  the QHC21 equation of state \cite{Kojo2022},  features a baryon chemical potential that is compatible within the 68\% credibility band of our inferred $\mu_B$.  While this data-inferred $\mu_B$  
cannot identify the detailed composition of species carrying baryon number,  it can nonetheless serve as a useful benchmark for testing models of the matter inside neutron star cores. 

(iv) Our updated inference results confirm the trend of the trace anomaly measure $\Delta = 1/3 - P/\varepsilon$  turning negative at energy densities above 500 MeV/fm$^{3}$,  corresponding to baryon densities $n_B > 3\,n_0$.  With inclusion of both the new J0437 NICER data and the black widow J0952-0607 mass,  the empirical evidence for $\Delta < 0$ resulting from a Bayes factor analysis (\ref{eq:bayes}) increases to the category {\it strong} throughout the range of energy densities relevant for heavy neutron stars.  Even omitting the J0952-0607 mass data  in the inference procedure,  there is still moderate evidence for negative trace anomaly measures inside neutron stars.  Hence, this empirical constraint is likely to relegate the quest for reaching the conformal limit in highly compressed cold matter to densities well beyond those realized in neutron star cores. 

As an outlook,  in view of the expected continuing supply of accurate neutron star data,  the present results can be taken as an illustrative example for the further progressive improvement of constraints on the EoS of cold dense matter. 

\section*{Acknowledgments}
Work partially supported by Deutsche Forschungsgemeinschaft (DFG) through the CRC110 "Symmetries and Emergence of Structure in QCD",  and by the DFG Excellence Cluster ORIGINS (EXC-2094).  One of the authors (W.W.) gratefully acknowledges the hospitality of the Yukawa Institute for Theoretical Physics at Kyoto University during the extended workshop HHIQCD2024 where this paper has been finalized.  He thanks Yuki Fujimoto,  Tetsuo Hatsuda,  Etsuko Itou,  Toru Kojo,  Jim Lattimer,  Owe Philipsen and Rob Pisarski for stimulating discussions.
\\
\\
%\appendix
%\section{EoS tables}

%\bibliography{library}

\end{document}